\documentclass[prb,twocolumn,showpacs,superscriptaddress]{revtex4}

\usepackage{graphicx}
\usepackage{dcolumn}
\usepackage{amsmath}
\usepackage{color}

\newcommand{\SNVO}{SrNi$_{2}$V$_2$O$_8$}
\newcommand{\PNVO}{PbNi$_{2}$V$_2$O$_8$}

\begin{document}
\title{Low-energy magnetic excitations in the quasi-one-dimensional spin-1 chain compound SrNi$_{2}$V$_2$O$_8$}

\author{Zhe~Wang}
\author{M.~Schmidt}
\affiliation{Experimental Physics V, Center for Electronic
Correlations and Magnetism, Institute of Physics, University of Augsburg, D-86135 Augsburg, Germany}

\author{A. K. Bera}
\author{A. T. M. N. Islam}
\affiliation{Helmholtz-Zentrum Berlin f\"{u}r Materialien und
Energie, D-14109 Berlin, Germany}

\author{B. Lake}
\affiliation{Helmholtz-Zentrum Berlin f\"{u}r Materialien und
Energie, D-14109 Berlin, Germany}

\affiliation{Institut f\"{u}r Festk\"{o}rperphysik, Technische
Universit\"{a}t Berlin, D-10623 Berlin, Germany}

\author{A.~Loidl}
\author{J.~Deisenhofer}
\affiliation{Experimental Physics V, Center for Electronic
Correlations and Magnetism, Institute of Physics, University of Augsburg, D-86135 Augsburg, Germany}

\date{\today}

\begin{abstract}
Multi-frequency electron spin resonance (ESR) transmission spectra have been measured as function of temperature and magnetic field on single crystals of the quasi-one-dimensional spin-1 chain compound \SNVO~in the GHz frequency range. Magnetic resonance modes above 50~K have been observed with an effective \emph{g}-factor of 2.24 at 100~K. Below 30~K, intra-triplet excitations have been observed in the ESR spectra, which reveal the presence of single-ion anisotropy with $D=-0.29$~meV.

\end{abstract}



\maketitle

\section{Introduction}
The one-dimensional spin-1 system with nearest-neighbor Heisenberg antiferromagnetic exchange $J$ exhibits a magnetic excitation gap, \emph{i.e.} the Haldane gap, $\Delta=0.41J$ between a spin-singlet ground state and a spin-triplet excited state.\cite{Haldane83,Haldane83a,White929308} The effects of interchain exchange interactions  $J_\perp$ and single-ion anisotropy $D$  on the ground state and excitation gap have been extensively studied in various theoretical approaches.\cite{Botet83,Pakinson85,Nightingale86,Sakai90,Golinelli93} An ideal one-dimensional spin-1 system does not exhibit long-range order even at $T=0$~K due to strong spin fluctuations, but in real systems long-range order can form at finite temperature when sufficiently large interchain exchange interactions exist. The magnetic excitation gap can be reduced or completely suppressed with the presence of interchain exchange interaction and single-ion anisotropy.\cite{Botet83,Pakinson85,Nightingale86,Sakai90,Golinelli93}

\begin{figure}[t]
\centering
\includegraphics[width=75mm,clip]{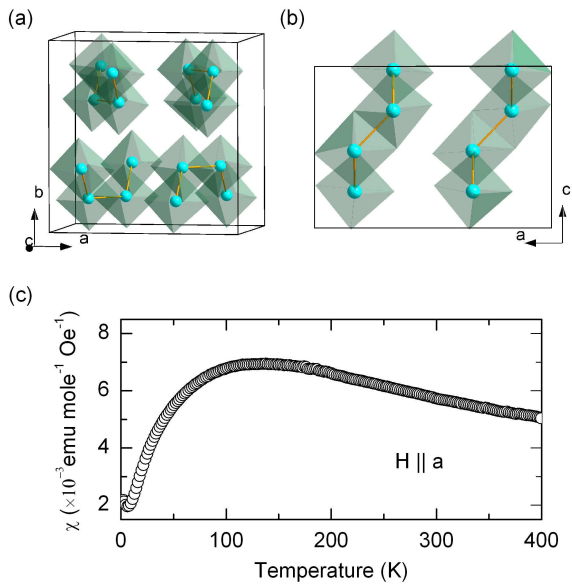}
\vspace{2mm} \caption[]{\label{Fig:Structure} (Color
online) (a) Edge-shared NiO$_6$ octahedra surrounding the fourfold screw axis forming spin chains along the c-axis in a unit cell of \SNVO. Ni ions are shown by the spheres. (b) Two spin chains along the c-axis.  (c) Temperature dependence of the magnetic susceptibility measured at 1~T with the magnetic field parallel to the \emph{a}-axis.}
\end{figure}

The isotructural spin-chain compounds \emph{A}Ni$_{2}$V$_2$O$_8$ (\emph{A}~=  Pb and Sr) based on Ni$^{2+}$ ions with $S=1$ have attracted considerable attention\cite{Uchiyama99,Zheludev00,Sirnov02,Pahari06,Sirnov08,He2008,Bera12} because the excitation gap ($0.23J$ with $J=8.2$~meV for \emph{A} =  Pb,\cite{Uchiyama99} and $0.25J$ with $J=9.2$~meV for \emph{A} =  Sr\cite{Bera12}) is significantly smaller than the expected $\Delta=0.41J$. This indicates that these compounds are close to a phase boundary in the $D$--$J_\perp$ phase diagram.\cite{Sakai90} \emph{A}Ni$_{2}$V$_2$O$_8$ crystallizes in tetragonal symmetry with space group $I4_1cd$, where the edge-shared NiO$_6$ octahedra surrounding the fourfold screw axis form screw chains along the \emph{c}-axis [Figs.~\ref{Fig:Structure}(a) and (b)].\cite{Bera12} By slightly doping at the Ni-site or the \emph{A}-site with nonmagnetic ions, the ground state changes from a disordered spin-liquid state to a N\'{e}el ordered state.\cite{Uchiyama99,Sirnov02,He2008} The nonmagnetic substitutions at the Ni-sites introduce free spins at the ends of spin chains. The coupling of these free spins due to inter-chain interaction can lead to the long-range magnetic order.\cite{Uchiyama99} For the \emph{A}-site doping the disorder-to-order transition has been ascribed to the modification on the interchain exchange interactions by substitutions.\cite{He2008}

According to the report on powder samples,\cite{Uchiyama99,Zheludev00} the ground state of \SNVO~is suggested as three-dimensional ordered state below 7~K due to Dzyaloshinskii-Moriya interaction, which is different from the spin-liquid state of \PNVO. Recently, nuclear magnetic resonance measurements performed on powder samples indicate that \SNVO~is disordered down to 3.75~K.\cite{Pahari06} The absence of an anomaly in the specific heat also rules out the long-range ordering in \SNVO~above 2~K.\cite{He2008} The magnetic susceptibility of \emph{A}Ni$_{2}$V$_2$O$_8$ increases with decreasing temperature and exhibits a broad maximum around 130~K [see Fig.~\ref{Fig:Structure}(c) for \SNVO]. At low temperatures below 10~K, an upturn of the susceptibility has been observed due to the Curie-Weiss contribution of magnetic impurities in the sample.\cite{Uchiyama99,Pahari06,He2008,Bera12} The susceptibility curve does not exhibit sharp anomalies down to 2~K, indicating the absence of long-range magnetic ordering.\cite{Pahari06,He2008,Bera12}


Based on a perturbative approach to describe inelastic neutron scattering results,\cite{Golinelli93} the single-ion anisotropy in \PNVO~and \SNVO~has been estimated as $-0.45$ and $-0.56$~meV, respectively.\cite{Zheludev00} An ESR study reported a smaller value of about $-0.31$~meV for \PNVO,\cite{Sirnov08} in agreement with an estimate from magnetization measurements.


In this work, we investigate the excitation spectra of \SNVO~single crystals in the GHz frequency range by multi-frequency ESR transmission spectroscopy. We have observed the intra-triplet excitations in the ESR spectra. The analysis of the field dependence of the intra-triplet excitation spectra reveals the existence of significant single-ion anisotropy in \SNVO.


\section{Experimental details}

Single crystals of \SNVO~were grown by the traveling solvent floating zone technique, the details of which will be published elsewhere.\cite{Bera2012Neutron} Magnetic susceptibility was measured with a constant magnetic field of 1~T parallel to the \emph{a}-axis in the warming cycle from 2 to 400~K using a Quantum Design physical properties measurement system. The single crystals for optical measurements were oriented by Laue diffraction, cut perpendicular to the \emph{a}-axis with dimension of $2\times 2\times 1$~mm$^{3}$ and perpendicular to the \emph{c}-axis with $2\times 2\times 0.7$~mm$^{3}$, and polished. Multi-frequency transmission experiments were performed for 2~$<T<$~210~K with an applied magnetic field varying up to 7~T in Voigt configuration for the external magnetic field \textbf{H} parallel to the \emph{a}-axis ($\textbf{H} \parallel\,a$) and the \emph{c}-axis ($\textbf{H} \parallel\,c$) with backward-wave oscillators covering the frequency range 125 - 500~GHz and a magneto-optical cryostat (Oxford/Spectromag).

\section{Experimental results and discussions}

\begin{figure}[t]
\centering
\includegraphics[width=85mm,clip]{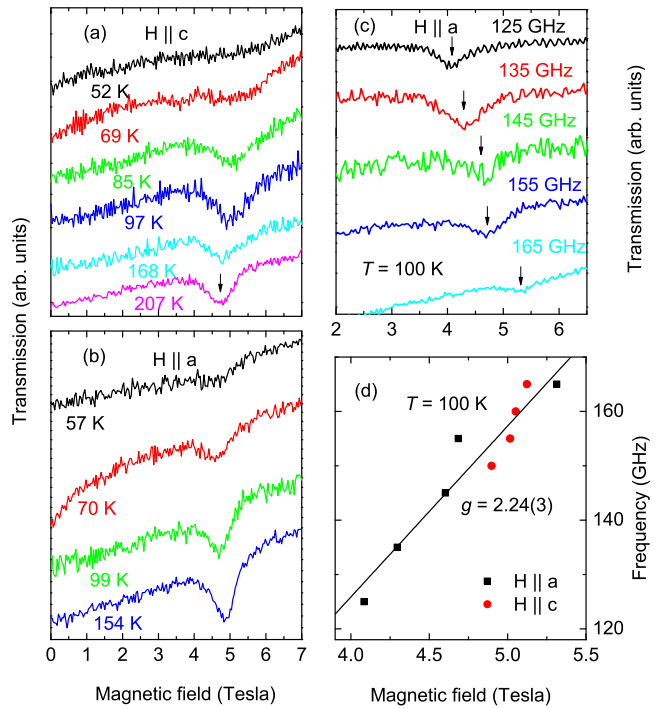}
\vspace{2mm} \caption[]{\label{Fig:F155GHzHochT} (Color
online) Transmission spectra at 155~GHz measured for various temperature with the magnetic field (a) parallel to the c-axis and (b) parallel to the a-axis. (c) Transmission spectra at various frequencies measured for 100~K with $\textbf{H} \parallel a$. (d) Resonance frequency as a function of resonance field at 100~K. Linear fit gives a \emph{g}-factor of 2.24(3).}
\end{figure}

Figures~\ref{Fig:F155GHzHochT}(a) and (b) show the transmission as a function of magnetic field measured at 155~GHz for various temperatures above 50~K for $\textbf{H} \parallel c$ and $\textbf{H} \parallel a$, respectively. At 207~K, a resonance mode is observed at 4.73~T for $\textbf{H}\parallel c$ corresponding to an exchange-narrowed Ni$^{2+}$ ESR signal, as marked by the arrow in Fig.~\ref{Fig:F155GHzHochT}(a). The resonance field increases with decreasing temperature and reaches 5.27~T at 69~K. Below 50~K this mode cannot be resolved. As shown in Fig.~\ref{Fig:F155GHzHochT}(b), the resonance mode can be also observed above 50~K for $\textbf{H} \parallel a$. The temperature dependence is consistent with the isostructural compound \PNVO.\cite{Sirnov08} The transmission spectra as a function of magnetic field corresponding to several frequencies have been measured for $\textbf{H} \parallel\,c$ and for $\textbf{H} \parallel\,a$. Fig.~\ref{Fig:F155GHzHochT}(c) shows the transmission spectra at 100~K at various frequencies with $\textbf{H} \parallel a$, which reveal resonance modes as marked by the arrows. The resonance modes have also been observed in the transmission spectra for $\textbf{H} \parallel c$ (not shown). The observed resonance frequencies versus resonance magnetic fields at 100~K are shown in Fig.~\ref{Fig:F155GHzHochT}(d). The field dependence of resonance frequencies exhibits a linear relation corresponding to the paramagnetic resonance. The experimental data can be fitted by a linear function through the origin resulting in a \emph{g}-factor of 2.24(3) for both field orientations. The line in Fig.~\ref{Fig:F155GHzHochT}(d) shows the fit result that is in good agreement with the experimental data. The obtained \emph{g}-factor is consistent with 2.23 determined in \PNVO,\cite{Sirnov08,Sirnov02} which is a typical value for Ni$^{2+}$ ions in octahedral crystal fields.\cite{Abragam1970}

The intensity of the ESR modes shown in Figs.~\ref{Fig:F155GHzHochT}(a) and (b) increases slightly with decreasing temperature and then decreases significantly below 100~K. This is consistent with the temperature dependence of the magnetic susceptibility, which exhibits a broad maximum around 130~K [see Fig.~\ref{Fig:Structure}(c)]. At low temperatures, the ESR modes are quite broad, have low intensity, and cannot be tracked below 50~K.

\begin{figure}[t]
\centering
\includegraphics[width=82mm,clip]{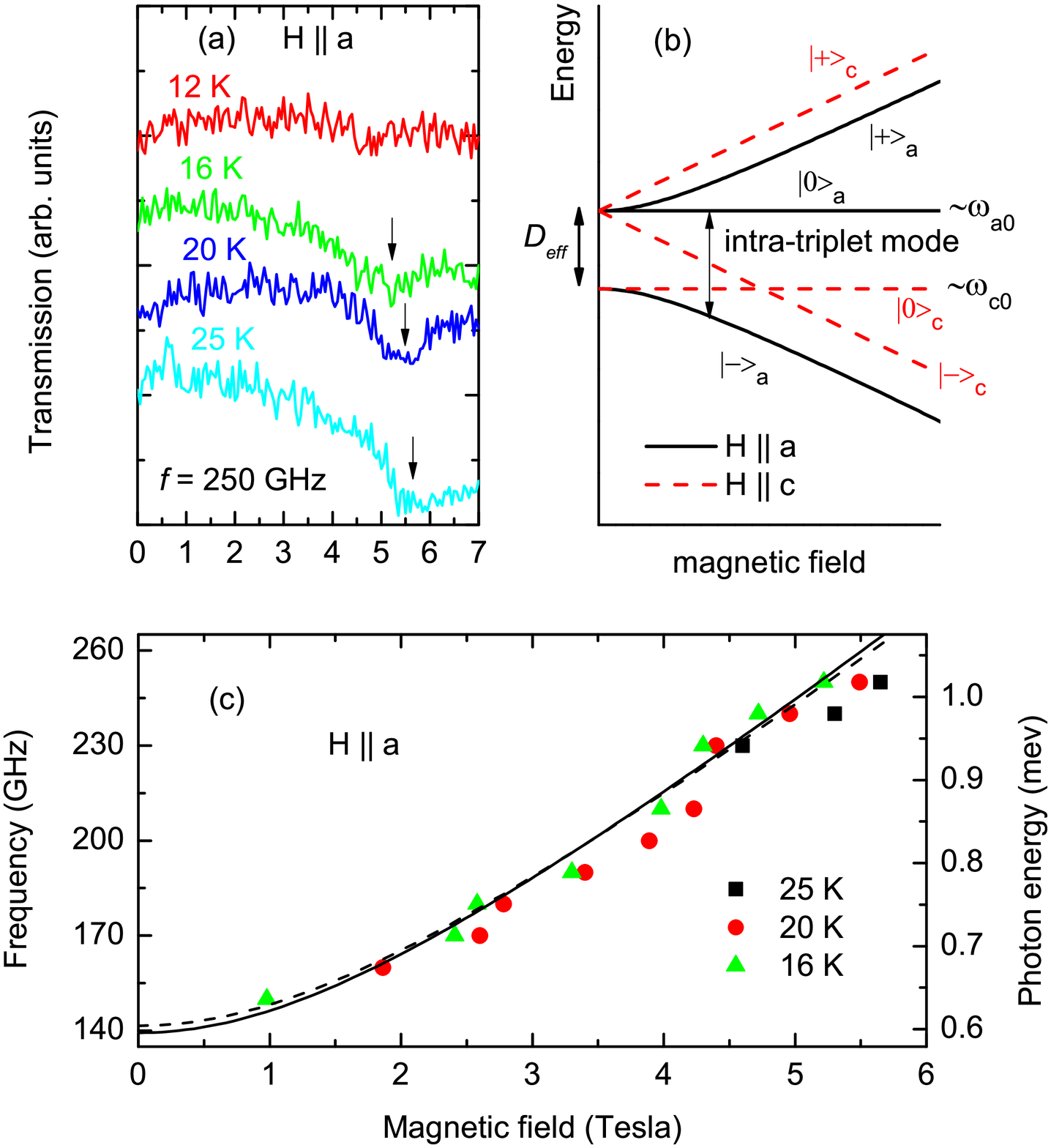}
\vspace{2mm} \caption[]{\label{Fig:F250GHzTiefT} (Color online) (a) Transmission spectra of 250~GHz measured at various temperatures with $\textbf{H} \parallel\,a$. (b) Zeeman splitting scheme of spin-triplet states with single-ion anisotropy $D<0$ for $\textbf{H} \parallel\,a$ (solid lines) and $\textbf{H} \parallel\,c$ (dashed lines)  according to the macroscopic field theory.\cite{Farutin07,Sirnov08} (c) Resonance frequencies versus resonance magnetic fields determined from various spectra measured at various temperatures with $\textbf{H} \parallel\,a$. The solid line and dashed line are fits of the experimental data obtained at 16~K according to the macroscopic field theory [Eq.~(\ref{Eq:MFT})] and the perturbation approach [Eq.~(\ref{Eq:PT})], respectively.}

\end{figure}

Figure~\ref{Fig:F250GHzTiefT}(a) shows the transmission spectra at 250~GHz measured below 30~K for $\textbf{H} \parallel a$.
A new absorption mode around 5.5~T can be seen at 25~K. The resonance field of this mode decreases slightly with decreasing temperature.
This mode cannot be resolved when the temperature is below 12~K. We assign this mode to the intra-triplet resonance mode of Ni$^{2+}$ ions [see Fig.~\ref{Fig:F250GHzTiefT}(b)]. At very low temperatures (below 12~K), the lower-lying triplet branch becomes much less populated due to the magnetic gap,
therefore the intra-triplet excitations cannot be observed.
Below 30~K, the splitting of spin-triplet state has also been observed in the inelastic neutron scattering experiments.\cite{Bera2012Neutron}
We did not observe any mode in the transmission spectra at 250~GHz measured in the same temperature range for $\textbf{H} \parallel c$ (not shown), which indicates that a significant single-ion anisotropy exists in this compound.\cite{Golinelli93,Sirnov08}

The single-ion anisotropy contributes to the Hamiltonian in the form $D(S_i^z)^2$ with the anisotropy constant $D$, which gives rise to the zero-field splitting of the spin-triplet state. Theoretically, an effective anisotropy $D_{\text{\emph{eff}}}\equiv-1.98D$ corresponding to the intra-triplet splitting in spin-1 systems has been studied\cite{Golinelli93} and applied to explain the results of \PNVO.\cite{Sirnov08} With the presence of single-ion anisotropy, the dependence of triplet states on the external magnetic field is quite different for $\textbf{H} \parallel\,c$ and for $\textbf{H} \parallel\,a$ [see Fig.~\ref{Fig:F250GHzTiefT}(b) for $D<0$ (easy-axis type)].\cite{Golinelli93,Sirnov08}
For $D_{\text{\emph{eff}}}>0$, we use $|-\rangle_a$ to denote the lower-lying state for $\textbf{H} \parallel a$, which has lowering energy with increasing magnetic field. The state independent on magnetic field for $\textbf{H} \parallel c$ is denoted by $|0\rangle_c$, which has lowest energy when the magnetic field is smaller than 4.4~T (estimated from $D_{\text{\emph{eff}}}$, see below). Consequently, the most probable intra-triplet excitations will be from $|-\rangle_a$ to $|0\rangle_a$ for $\textbf{H} \parallel a$ and from $|0\rangle_c$ to $|\pm\rangle_c$ for $\textbf{H} \parallel c$. With increasing magnetic field, the absorption mode corresponding to $|-\rangle_a\rightarrow|0\rangle_a$ for $\textbf{H} \parallel a$ will have more intensity and shifts to higher frequency due to the lowering of the energy of the $|-\rangle_a$ state by the Zeeman contribution. However, the intensity corresponding to  $|0\rangle_c\rightarrow|\pm\rangle_c$ is rather constant for $\textbf{H} \parallel c$ since the population of the $|0\rangle_c$ state will not be altered by the magnetic field.

The ESR results indicate that the single ion anisotropy is of the single-axis type \emph{i.e.}, $D<0$, otherwise the excitations should be observed for $\textbf{H} \parallel c$. The resonance frequencies versus resonance fields of the observed modes for $\textbf{H} \parallel a$ are plotted in Fig.~\ref{Fig:F250GHzTiefT}(c). Below 1~T, we cannot resolve any modes due to the low population of the $|-\rangle_a$ state. This is consistent with the fact that no modes corresponding to $|0\rangle_c\rightarrow|\pm\rangle_c$ can be resolved from the spectra for $\textbf{H} \|\,c$, which should have lower intensity compared to $|-\rangle_a\rightarrow|0\rangle_a$ for $\textbf{H} \|\,a$ in a finite magnetic field [see Fig.~\ref{Fig:F250GHzTiefT}(b)]. According to the macroscopic field theory approach,\cite{Farutin07,Sirnov08} the field dependence of the frequency difference between the states $|0\rangle_a$ and $|-\rangle_a$ for $\textbf{H} \parallel a$ can be described by
\begin{equation}\label{Eq:MFT}
\omega(H)=\sqrt{A+B}-\sqrt{(\gamma H)^2+A-\sqrt{B^2+4A(\gamma H)^2}},
\end{equation}
where $\gamma= g\mu_B$ with $\mu_B$ the Bohr magneton, $A\equiv(\omega_{a0}^2+\omega_{c0}^2)/2$ and $B\equiv(\omega_{a0}^2-\omega_{c0}^2)/2$ with $\omega_{a0}$ the eigenfrequency of the $|0\rangle_a$ state for $\textbf{H} \parallel a$ and $\omega_{c0}$ the eigenfrequency of the $|0\rangle_c$ state for $\textbf{H} \parallel c$ [see Fig.~\ref{Fig:F250GHzTiefT}(b)]. Using the perturbation theory,\cite{Golinelli93} the field dependence has been shown to follow the expression
\begin{equation}\label{Eq:PT}
\omega(H)=\frac{1}{2} \left[ \omega_{a0}-\omega_{c0} + \sqrt{(\omega_{a0}-\omega_{c0})^2 +4(\gamma H)^2} \right].
\end{equation}

As shown by the solid line and dashed line in Fig.~\ref{Fig:F250GHzTiefT}(c), respectively, both fits to the experimental data at 16~K with Eq.~(\ref{Eq:MFT}) and Eq.~(\ref{Eq:PT}) provide a good description of the experimental results, where we use the experimental \emph{g}-factor of 2.24. The fits according to the macroscopic field theory [Eq.~(\ref{Eq:MFT})] and perturbative theory [Eq.~(\ref{Eq:PT})] determine the intra-triplet splitting $\omega_{a0}-\omega_{c0}$ of 139~GHz and 141~GHz at zero magnetic field, respectively, which agree well with each other within the uncertainty. The intra-triplet splitting corresponds to $D_{\text{\emph{eff}}}=0.57$~meV. The single-ion anisotropy $D$ can be consequently estimated as $-0.29$~meV, which is close to $-0.4$~meV determined at 16~K by the inelastic neutron scattering experiments.\cite{Bera2012Neutron}






\section{Conclusion}

Multi-frequency electron spin resonance transmission spectra have been measured as a function of temperature on single crystalline spin-1 chain compound \SNVO~with the external magnetic field parallel and perpendicular to the spin-chain direction.
Above 50~K, the exchange-narrowed Ni$^{2+}$ paramagnetic resonance lines have been observed,
which determines a \emph{g}-factor of 2.24(3) at 100~K. Between 10 and 30~K,
intra-triplet resonance lines are observed for $\mathbf{H}\parallel a$ but not for $\mathbf{H}\parallel c$ in the same frequency range,
indicating the existence of single-ion anisotropy. The single-ion anisotropy is estimated as $-0.29$~meV from the resonance modes observed for $\mathbf{H}\parallel a$.

\begin{acknowledgments}
We acknowledge partial support by the Deutsche Forschungsgemeinschaft via TRR 80
(Augsburg-Munich) and Project DE 1762/2-1.
\end{acknowledgments}

\end{document}